\documentclass[prl,aps,superscriptaddress,twocolumn,reprint,longbibliography,floatfix,noeprint]{revtex4-2}
\usepackage{amsmath,amssymb,amsthm,mathtools}
\usepackage{bm}
\usepackage{graphicx}
\usepackage[colorlinks=true,linkcolor=blue,citecolor=blue,urlcolor=blue]{hyperref}

\theoremstyle{plain}
\newtheorem{theorem}{Theorem}
\newtheorem{proposition}{Proposition}
\newtheorem{corollary}{Corollary}
\newtheorem{lemma}{Lemma}

\newcommand{\Fp}{\mathbb{F}_{p}}
\newcommand{\Vn}{V_{n}}
\newcommand{\Vlog}{V_{\mathrm{log}}}
\newcommand{\Sp}{\operatorname{Sp}}
\newcommand{\SL}{\operatorname{SL}}
\newcommand{\Sym}{\operatorname{Sym}}
\newcommand{\MonSp}{\operatorname{MonSp}}
\newcommand{\Autmon}{\operatorname{Aut}_{\mathrm{mon}}}
\newcommand{\Id}{\mathrm{Id}}
\newcommand{\ord}{\operatorname{ord}}
\newcommand{\Ann}{\operatorname{Ann}}
\newcommand{\SUM}{\mathrm{SUM}}

\begin{document}

\title{No-Go Theorems for Transversal and Automorphism Clifford Gates on Prime-Dimensional Qudit Stabilizer Codes}

\author{Yuchen Zhan}\thanks{These authors contributed equally to this work.}
\affiliation{Center for Quantum Information, IIIS, Tsinghua University, Beijing 100084, People's Republic of China}

\author{Qi Ye}\thanks{These authors contributed equally to this work.}
\affiliation{Center for Quantum Information, IIIS, Tsinghua University, Beijing 100084, People's Republic of China}

\author{Dong-Ling Deng}
\email{dldeng@tsinghua.edu.cn}
\affiliation{Center for Quantum Information, IIIS, Tsinghua University, Beijing 100084, People's Republic of China}
\affiliation{Shanghai Qi Zhi Institute, Shanghai 200232, China}
\affiliation{Hefei National Laboratory, Hefei 230088, China}

% Abstract: <= 600 characters including spaces (PRL).
\begin{abstract}
Logical Clifford gates are a basic primitive of fault-tolerant quantum computation. For stabilizer codes on qudits of prime dimension $p$ that encode $k\geq 2$ logical qudits, we prove that Clifford gadgets that are transversal with respect to a fixed partition of the physical qudits into blocks of fewer than $k$ qudits cannot realize the full logical Clifford group, and that this block size is tight. Code automorphisms, which combine single-qudit Clifford gates with qudit permutations, also fail for every $p$. Both obstructions come from prime-order logical Clifford elements that these gadgets cannot reach.
\end{abstract}

\date{\today}
\maketitle

Quantum error correction protects logical information by encoding it into a larger physical system~\cite{shorSchemeReducingDecoherence1995,calderbankGoodQuantumErrorcorrecting1996,steaneErrorCorrectingCodes1996,gottesmanStabilizerCodesQuantum1997}.
A fault-tolerant quantum computer must also manipulate the encoded information without letting errors spread uncontrollably.
Clifford gates are central to this task, because they map Pauli errors to Pauli errors and, supplemented by magic states, yield universal fault-tolerant computation~\cite{bravyiUniversalQuantumComputation2005}.
The simplest fault-tolerant implementations are transversal gates, which act independently on each physical qudit.
For any code that detects single-site errors, however, transversal gates cannot form a universal gate set~\cite{eastinRestrictionsTransversalEncoded2009,zengTransversalityUniversalityAdditive2011}, and related constraints restrict more general local gadgets~\cite{bravyiClassificationTopologicallyProtected2013,jochym-oconnorDisjointnessStabilizerCodes2018}.

High-rate codes, such as quantum low-density parity-check codes, encode many logical systems in a single code block~\cite{bravyiHighthresholdLowoverheadFaulttolerant2024}.
For such codes, a central question is how much of the logical Clifford group can be reached with cheap physical gadgets.
Two extensions of transversality are being actively explored.
Code automorphisms combine single-site Clifford gates with permutations of the physical sites~\cite{sayginelFaultTolerantLogicalClifford2025,kohAchievingLimitsAutomorphism2026}, while fold-transversal gadgets also allow two-site Clifford gates on a fixed pairing of sites~\cite{breuckmannFoldTransversalCliffordGates2024,eberhardtLogicalOperatorsFoldTransversal2025}.
For qubits, Chakraborty and Gottesman proved that code automorphisms cannot realize the full logical Clifford group on two or more logical qubits, and that fold-transversal gadgets cannot do so on three or more~\cite{chakrabortyNoGoTheoremFault2026}.
More generally, they showed that $k$ logical qubits require $k$-fold transversal gadgets.
Related obstructions constrain addressable logical gates~\cite{guyotAddressabilityProblemCSS2026} and permutation-based logical circuits~\cite{morrisConstraintsPhantomCodes2026}, and for $k\geq 3$ the largest logical group reachable by qubit automorphisms has recently been identified~\cite{kohAchievingLimitsAutomorphism2026}.

Many quantum platforms, such as trapped ions and superconducting cavities, naturally provide more than two levels per physical carrier.
Using these additional levels can make gates and algorithms more efficient, and universal qudit processors as well as qudit error correction beyond break-even have recently been demonstrated~\cite{ringbauerUniversalQuditQuantum2022,brockQuantumErrorCorrection2025}.
On the theory side, qudit stabilizer codes share the stabilizer formalism of qubit codes~\cite{gottesmanFaultTolerantQuantumComputation1999,ketkarNonbinaryStabilizerCodes2006}, and they enable magic-state distillation schemes whose thresholds and yields can exceed those of their qubit counterparts and improve with the dimension~\cite{campbellMagicStateDistillationAll2012,campbellEnhancedFaultTolerantQuantum2014}.
Whether the qubit no-go theorems extend to qudits was posed as an open question in Ref.~\cite{chakrabortyNoGoTheoremFault2026}.
This generalization is not straightforward, because the qubit argument for code automorphisms relies on an explicit two-qubit Bell gate of order $5$, together with the fact that $5$ does not divide $6$, the order of the single-qubit Clifford group modulo Pauli operators and phases~\cite{chakrabortyNoGoTheoremFault2026}.
For odd $p$, both the relevant primes and the group orders depend on $p$, so a new argument that works uniformly in $p$ is needed.

In this paper, we prove both no-go theorems for stabilizer codes in every prime dimension $p$.
Theorem~\ref{thm:block} shows that Clifford gadgets that are transversal with respect to a fixed partition into blocks of fewer than $k$ qudits cannot realize the full logical Clifford group on $k$ logical qudits, and Proposition~\ref{prop:tight} shows that this threshold is sharp.
Theorem~\ref{thm:mon} shows that code automorphisms fail for every prime $p$ and every $k\geq 2$.
Its proof is geometric and uniform in $p$.
In place of the explicit qubit gate, it uses a logical element whose order is an odd prime $\ell$ dividing $p^{2}+1$ and which preserves no maximal commuting set of logical Pauli operators.
For $p=2$, the argument recovers the theorem of Ref.~\cite{chakrabortyNoGoTheoremFault2026}.
We also derive a consequence for addressable logical gates.
\section*{Setup}
Let $p$ be prime and $\omega=e^{2\pi i/p}$.
On a single qudit, the generalized Pauli operators act as $X|j\rangle=|j+1\rangle$ and $Z|j\rangle=\omega^{j}|j\rangle$ for $j$ in $\Fp$, the integers modulo $p$.
The $n$-qudit Pauli group $\mathcal{P}_n$ is generated by these operators on each qudit together with the phases $\omega^{j}$, and also $i$ when $p=2$.
The Clifford group $\mathcal{C}_n$ consists of the unitaries $U$ with $U\mathcal{P}_nU^\dagger=\mathcal{P}_n$.
It is generated, up to phases, by the qudit analogs of the Hadamard, $S$, and CNOT gates~\cite{farinholtIdealCharacterizationClifford2014,hostensStabilizerStatesClifford2005}.
These are the Fourier gate $F|j\rangle=p^{-1/2}\sum_{m}\omega^{jm}|m\rangle$, the phase gate $P|j\rangle=\omega^{j(j-1)/2}|j\rangle$ (with $P=\mathrm{diag}(1,i)$ for $p=2$), and the two-qudit gate $\SUM|j,m\rangle=|j,j+m\rangle$.
An $[[n,k]]_p$ stabilizer code is the joint $+1$ eigenspace of an abelian subgroup $S\subseteq\mathcal{P}_n$ that contains no nontrivial phase and has $p^{n-k}$ elements, and it encodes $k$ logical qudits~\cite{gottesmanFaultTolerantQuantumComputation1999,ketkarNonbinaryStabilizerCodes2006}.
A Clifford gadget is a Clifford unitary on the physical qudits that maps the code space to itself, or equivalently satisfies $USU^\dagger=S$.
After fixing an encoding of the code space as $k$ qudits, its restriction to the code space is a Clifford unitary on the $k$ logical qudits, which we call its logical action.
A group of gadgets realizes the full logical Clifford group if every logical Clifford unitary agrees, up to a logical Pauli operator and a phase, with the logical action of some gadget in the group.
This Pauli freedom is harmless, because logical Pauli operators are products of single-qudit Pauli operators and belong to both architectures below.

\section*{Main results}
\textit{Transversal gadgets with blocks.---}Following Refs.~\cite{chakrabortyNoGoTheoremFault2026,jochym-oconnorDisjointnessStabilizerCodes2018}, we fix a partition of the physical qudits into blocks $B_1,\ldots,B_m$.
A gadget is transversal with respect to this partition if it is a tensor product of Clifford unitaries, each acting on the qudits of a single block.
Such gadgets form a group, and they are called $t$-fold transversal when every block contains at most $t$ qudits.
Ordinary transversal gates correspond to $t=1$, and fold-transversal gates on a fixed pairing of qudits have $t=2$.

\begin{theorem}[Block no-go]\label{thm:block}
Let $k\geq 2$, and fix a partition of the physical qudits of an $[[n,k]]_p$ stabilizer code into blocks of fewer than $k$ qudits each.
Then the gadgets that are transversal with respect to this partition do not realize the full logical Clifford group.
\end{theorem}

In particular, transversal gadgets fail whenever $k\geq 2$, and gadgets on a fixed pairing of qudits fail whenever $k\geq 3$.
The obstruction is arithmetic and follows the qubit argument of Ref.~\cite{chakrabortyNoGoTheoremFault2026}.
Modulo Pauli operators and phases, the Clifford group on $s$ qudits is a finite group of order $p^{s^2}\prod_{i=1}^{s}(p^{2i}-1)$~\cite{hostensStabilizerStatesClifford2005,wilsonFiniteSimpleGroups2009}, and throughout, the order of a gate means its order in this group.
By the Bang--Zsigmondy theorem~\cite{zsigmondyZurTheoriePotenzreste1892}, some prime $\ell$ divides this order for $s=k$ but not for any $s<k$.
The logical Clifford group therefore contains a gate of order $\ell$.
On the other hand, no gadget built from blocks of fewer than $k$ qudits has an order divisible by $\ell$, and the order of a logical action divides the order of the gadget that induces it.

Theorem~\ref{thm:block} is sharp, because blocks of exactly $k$ qudits can suffice, as the following extension of the qubit example of Ref.~\cite{chakrabortyNoGoTheoremFault2026} to all primes shows.

\begin{proposition}[Tightness]\label{prop:tight}
For every prime $p$ and every $k\geq 1$, there is a stabilizer code that encodes $k$ logical qudits with distance at least $2$, together with a partition of its physical qudits into blocks of $k$ qudits, such that the gadgets transversal with respect to this partition realize the full logical Clifford group.
\end{proposition}

For odd $p$, the code consists of $k$ copies of a length-$p$ quantum Reed--Solomon code with parameters $[[p,1,(p+1)/2]]_p$, closely related to the quantum Reed--Solomon and polynomial codes of Refs.~\cite{grasslQuantumReedSolomon1999,aharonovFaultTolerantQuantumComputation2008}, and each block collects the $i$th qudit of every copy.
Transversal Fourier, phase, and $\SUM$ gates then generate the full logical Clifford group.
Theorem~\ref{thm:block} relies on all gadgets sharing a single partition, and recent constructions that combine several partitions obtain the full logical Clifford group and even universal gate sets~\cite{tansuwannontConstructionFullLogical2026,duaQubitCSSCodes2026}.

\textit{Code automorphisms.---}A code automorphism is a gadget of the form $(V_1\otimes\cdots\otimes V_n)P_\pi$, where each $V_i$ is a single-qudit Clifford unitary and $P_\pi$ permutes the physical qudits according to a permutation $\pi$.
When the permutation is implemented by relabeling the qudits, such gadgets do not spread errors within the code block~\cite{chakrabortyNoGoTheoremFault2026}, and for a given code they form a group.

\begin{theorem}[Automorphism no-go]\label{thm:mon}
For every $[[n,k]]_p$ stabilizer code with $k\geq 2$, the code automorphisms do not realize the full logical Clifford group.
\end{theorem}

The order argument does not apply here, because permutations of many qudits can have arbitrarily large prime orders, so the obstruction must be geometric.
A maximal commuting set is a maximal set of mutually commuting Pauli operators, taken modulo phases, such as the set of all $Z$-type operators.
We call it a product set if it is generated by one nontrivial single-qudit Pauli operator on each qudit.
The key ingredient is an odd prime $\ell$ dividing $p^2+1$, for example $\ell=5$ when $p=2$ or $p=3$.
Such an $\ell$ never divides $p(p^2-1)$, the order of the single-qudit Clifford group modulo Pauli operators and phases, so it plays the role that the prime $5$ plays for qubits.
Every code automorphism whose order is a power of $\ell$ preserves some product set, and hence a maximal commuting set of logical Pauli operators.
In contrast, the logical Clifford group contains a gate $g$ of order $\ell$ that preserves no maximal commuting set of logical Pauli operators.
Any automorphism inducing $g$ has a power whose order is a power of $\ell$ and whose logical action, a power of $g$, still preserves no such set.

The restriction to automorphisms of $\ell$-power order is essential.
A general automorphism need not preserve any maximal commuting set of logical Pauli operators, as an order-$3$ single-qubit Clifford gate on one qubit of the trivial $[[2,2]]_2$ code shows.
For $p=2$, the prime $\ell=5$ matches the order of the Bell gate of Ref.~\cite{chakrabortyNoGoTheoremFault2026} modulo Pauli operators and phases, and the argument recovers their theorem without an explicit gate.

\textit{Addressable gates.---}An addressable logical gate acts on a chosen logical qudit or pair of logical qudits and trivially on all others.
Since the Fourier, phase, and $\SUM$ gates generate the Clifford group up to phases, addressable versions of these gates on all logical qudits realize the full logical Clifford group in the sense defined above.
The two theorems therefore have the following direct consequence.

\begin{corollary}[Addressable gates]\label{cor:address}
For $k\geq 2$, no $[[n,k]]_p$ stabilizer code implements all addressable logical Fourier, phase, and $\SUM$ gates by code automorphisms.
Moreover, for any fixed partition into blocks of fewer than $k$ qudits, no such code implements all of them by gadgets that are transversal for that partition.
\end{corollary}

This extends the addressability corollary of Ref.~\cite{chakrabortyNoGoTheoremFault2026} to all prime dimensions.
For qubit CSS codes, Guyot and Jaques proved stronger, gate-specific no-go results~\cite{guyotAddressabilityProblemCSS2026}.

\section*{Proofs}
We first translate the setup into the language of Pauli labels, and then prove Theorem~\ref{thm:block}, Proposition~\ref{prop:tight}, and Theorem~\ref{thm:mon} in turn.

\textit{Pauli labels and logical actions.---}Up to phases, $n$-qudit Pauli operators $W(a|b)=\bigotimes_{i}X^{a_i}Z^{b_i}$ are labeled by vectors $v=(a|b)$ in $\Vn=\Fp^{2n}$.
With the symplectic form $\langle (a|b),(a'|b')\rangle=a\cdot b'-b\cdot a'$, one has $W(v)W(w)=\omega^{-\langle v,w\rangle}W(w)W(v)$.
The commutation phase $\omega^{-\langle v,w\rangle}$ generalizes the sign $\pm1$ of the qubit case.
We write $\Sp(2n,p)$ for the group of $2n\times 2n$ matrices over $\Fp$ that preserve this form, and $\Sp(V)$ for the analogous group of any symplectic space $V$.
A Clifford unitary $U$ acts on labels by a symplectic matrix $\Gamma(U)$, defined by $UW(v)U^\dagger\propto W(\Gamma(U)v)$.
The map $\Gamma$ is onto $\Sp(2n,p)$, and its kernel consists of Pauli operators and phases~\cite{hostensStabilizerStatesClifford2005}.

The labels of the stabilizer $S$ form a subspace $L\subseteq\Vn$ of dimension $n-k$.
It is isotropic, meaning that the form vanishes on $L$, and its symplectic complement is $L^\perp=\{v\in\Vn:\langle v,s\rangle=0\text{ for all }s\in L\}$.
For $s\in L$, we write $g_s\in S$ for the unique element with label $s$, so that $s\mapsto g_s$ is an isomorphism from $(L,+)$ onto $S$.
Logical Pauli operators modulo phases correspond to the $2k$-dimensional symplectic space $\Vlog=L^\perp/L$~\cite{ketkarNonbinaryStabilizerCodes2006}.
Here $L^\perp$ collects the Pauli errors that are invisible to syndrome measurement, and quotienting by $L$ identifies errors that differ by a stabilizer.
If $M\in\Sp(\Vn)$ satisfies $M(L)=L$, it also preserves $L^\perp$ because it preserves the form, and hence induces $\rho_L(M)\in\Sp(\Vlog)$ via $[x]\mapsto[Mx]$.
Since $\rho_L$ is a homomorphism, the order of $\rho_L(M)$ divides the order of $M$.
For a gadget $U$, the map $\rho_L(\Gamma(U))$ describes how its logical action transforms logical Pauli operators.
Logical Clifford unitaries modulo logical Pauli operators and phases correspond one to one to elements of $\Sp(\Vlog)\cong\Sp(2k,p)$.
A group $\mathcal{G}$ of gadgets therefore realizes the full logical Clifford group exactly when $\rho_L(\Gamma(\mathcal{G}))=\Sp(\Vlog)$.

Gadgets that are transversal for a partition $\Pi=\{B_j\}$ have label maps in $G_\Pi=\bigoplus_j\Sp(2|B_j|,p)$.
Code automorphisms have label maps in the monomial group $\MonSp_n(p)=\Sp(2,p)^n\rtimes\Sym(n)$, where $\Sym(n)$ is the symmetric group.
This group permutes the label planes $E_i=\operatorname{span}\{x_i,z_i\}$ of the qudits, with $x_i=(e_i|0)$ and $z_i=(0|e_i)$, and acts on each plane by a one-qudit symplectic map.
The planes are mutually orthogonal and $\langle x_i,z_i\rangle=1$.
For a group $G\subseteq\Sp(2n,p)$, we write $G(L)=\{M\in G: M(L)=L\}$ and $\Autmon(L)=\{u\in\MonSp_n(p): u(L)=L\}$.
Every element of $G_\Pi$ or $\MonSp_n(p)$ is the label map of a Clifford unitary of the corresponding form.
The next lemma shows that, if this label map preserves $L$, a Pauli correction, which is a product of single-qudit Pauli operators, turns the unitary into a gadget.

\begin{lemma}[Pauli repair]\label{lem:pauli}
Let $U$ be a Clifford unitary with $M=\Gamma(U)$ and $M(L)=L$.
Then there is a Pauli operator $W$ such that $WUg_s(WU)^\dagger=g_{Ms}$ for all $s\in L$.
In particular, $WU$ preserves the code space and induces $\rho_L(M)$.
\end{lemma}

\begin{proof}
Write $Ug_sU^\dagger=\chi(s)\,g_{Ms}$ with a scalar $\chi(s)$.
Since $s\mapsto g_s$ is a homomorphism and $S$ is abelian, $\chi(s+t)=\chi(s)\chi(t)$.
Every $g\in S$ satisfies $g^p=I$, because $g^p$ is a scalar in $S$ and $S$ contains no nontrivial phase.
Conjugation preserves this property, so $\chi(s)^p=1$ and $\chi(s)=\omega^{f(s)}$ for a linear functional $f$ on $L$.
By nondegeneracy, there is $w\in\Vn$ with $\langle w,Ms\rangle=f(s)$ for all $s\in L$.
Then $W(w)g_{Ms}W(w)^\dagger=\omega^{-f(s)}g_{Ms}$, so $W=W(w)$ works.
\end{proof}

The gadgets of the two architectures thus have label maps exactly $G_\Pi(L)$ and $\Autmon(L)$.
Theorems~\ref{thm:block} and~\ref{thm:mon} are therefore equivalent to $\rho_L(G_\Pi(L))\neq\Sp(\Vlog)$ and $\rho_L(\Autmon(L))\neq\Sp(\Vlog)$, respectively.

\textit{Order obstruction.---}A primitive prime divisor of $p^r-1$ is a prime $\ell$ with $\ell\mid p^r-1$ and $\ell\nmid p^t-1$ for $1\leq t<r$.
We write $\ord_\ell(p)$ for the multiplicative order of $p$ modulo $\ell$, so a primitive prime divisor of $p^r-1$ has $\ord_\ell(p)=r$.
By the Bang--Zsigmondy theorem, such a prime exists for every $r>1$, except for $r=2$ with $p+1$ a power of $2$, and for $(p,r)=(2,6)$~\cite{zsigmondyZurTheoriePotenzreste1892}.

\begin{lemma}[Rank-$k$ prime]\label{lem:rank}
For $k\geq 2$, there is a prime $\ell$ with $\ell\mid\lvert\Sp(2k,p)\rvert$ and $\ell\nmid\lvert\Sp(2s,p)\rvert$ for all $1\leq s<k$.
\end{lemma}

\begin{proof}
If $k$ is odd, then $k\geq 3$, and we take a primitive prime divisor $\ell$ of $p^k-1$, so that $\ord_\ell(p)=k$.
If $k$ is even, we take a primitive prime divisor $\ell$ of $p^{2k}-1$.
The exceptions do not arise, because the exponent $k$ is odd in the first case, and in the second case $2k\geq 4$, while $(p,2k)=(2,6)$ would require $k=3$.
In both cases $\ell\neq p$ and $\ell\mid p^{2k}-1$, so $\ell\mid\lvert\Sp(2k,p)\rvert$.
If $\ell\mid\lvert\Sp(2s,p)\rvert$ for some $s<k$, then $\ell\mid p^{2i}-1$ for some $i\leq s$, so $\ord_\ell(p)$ divides $2i$.
For odd $k$ this forces $k\mid i$, and for even $k$ it forces $2k\mid 2i$.
Both contradict $i<k$.
\end{proof}

\begin{proof}[Proof of Theorem~\ref{thm:block}]
Take $\ell$ from Lemma~\ref{lem:rank}.
By Cauchy's theorem, $\Sp(\Vlog)\cong\Sp(2k,p)$ contains an element of order $\ell$.
For $M=\bigoplus_j M_j\in G_\Pi(L)$, each $\ord M_j$ divides $\lvert\Sp(2|B_j|,p)\rvert$ with $|B_j|<k$, so $\ell\nmid\ord M=\operatorname{lcm}_j\ord M_j$.
Since the order of $\rho_L(M)$ divides $\ord M$, the element $\rho_L(M)$ cannot have order $\ell$.
Hence $\rho_L(G_\Pi(L))\neq\Sp(\Vlog)$.
\end{proof}

\textit{Tightness construction.---}We now construct the codes and partitions of Proposition~\ref{prop:tight}.

\begin{proof}[Proof of Proposition~\ref{prop:tight}]
For $p=2$, $k$ copies of the Steane code~\cite{steaneErrorCorrectingCodes1996} work, as shown in Ref.~\cite{chakrabortyNoGoTheoremFault2026}.
For odd $p$, let $D\subset\Fp^{p}$ consist of the evaluations, at all $x\in\Fp$, of polynomials of degree less than $(p-1)/2$.
Since $\sum_{x\in\Fp}x^j=0$ for $0\leq j<p-1$ (with $0^0=1$), the evaluations of polynomials of degree at most $(p-1)/2$ lie in $D^\perp$.
Evaluation is injective on polynomials of degree less than $p$, so $\dim D=(p-1)/2$, and comparing dimensions shows that $D^\perp$ consists exactly of these evaluations.
Hence $D\subset D^\perp=D\oplus\Fp c$ with $c=(x^{(p-1)/2})_{x\in\Fp}$, and $c\cdot c=\sum_x x^{p-1}=-1$.
The CSS code with $L=(D|0)\oplus(0|D)$ encodes one qudit.
Its distance is at least $(p+1)/2$, the minimum distance of the extended Reed--Solomon code $D^\perp$, and equality holds because $D$ has the larger minimum distance $(p+3)/2$.
Its logical labels are $\bar X=[(c|0)]$ and $\bar Z=[(0|-c)]$, with $\langle\bar X,\bar Z\rangle=1$.

The transversal gates $F^{\otimes p}$ and $P^{\otimes p}$ act on labels as $(a|b)\mapsto(-b|a)$ and $(a|b)\mapsto(a|a+b)$, and both maps preserve $L$.
The first induces $\bar X\mapsto-\bar Z$ and $\bar Z\mapsto\bar X$, and the second induces $\bar X\mapsto\bar X-\bar Z$ and $\bar Z\mapsto\bar Z$.
These are the label maps of the logical gates $F^{-1}$ and $P^{-1}$.
In the basis $(\bar X,\bar Z)$ the second map is lower unitriangular, and its conjugate by the first is upper unitriangular.
Nontrivial lower and upper unitriangular matrices generate $\SL(2,p)=\Sp(2,p)$.

Now take $k$ copies and let block $B_i$ contain the $i$th qudit of every copy.
The transversal $\SUM$ gate from copy $\alpha$ to copy $\beta$ maps $(d|0)$ in copy $\alpha$ to $(d|0)$ in both copies, maps $(0|d)$ in copy $\beta$ to $-(0|d)$ in copy $\alpha$ plus $(0|d)$ in copy $\beta$, and fixes $(0|d)$ in copy $\alpha$ and $(d|0)$ in copy $\beta$.
It therefore preserves $L$ and induces the logical $\SUM$ gate, which maps $\bar X_\alpha\mapsto\bar X_\alpha+\bar X_\beta$ and $\bar Z_\beta\mapsto\bar Z_\beta-\bar Z_\alpha$ and fixes $\bar Z_\alpha$ and $\bar X_\beta$.
Single-qudit symplectic maps and $\SUM$ gates generate $\Sp(2k,p)$~\cite{farinholtIdealCharacterizationClifford2014,hostensStabilizerStatesClifford2005}.
All of these gadgets are $k$-fold transversal for this partition, and so are the Pauli corrections of Lemma~\ref{lem:pauli}.
\end{proof}

\textit{Lagrangian obstruction.---}In label language, a maximal commuting set is a Lagrangian, that is, a maximal isotropic subspace, or equivalently an isotropic subspace of half the dimension.
A product set is a coordinate Lagrangian $F=\bigoplus_i F_i$ built from lines $F_i\subset E_i$, and a maximal commuting set of logical Pauli operators is a Lagrangian of $\Vlog$.
We call an element $\ell$-primary if its order is a power of $\ell$.
The proof of Theorem~\ref{thm:mon} uses four lemmas.
The first fixes the prime $\ell$, the next two show that $\ell$-primary automorphisms preserve a logical Lagrangian, and the last constructs a logical element of order $\ell$ that preserves none.

\begin{lemma}[The prime $\ell$]\label{lem:ell}
There is an odd prime $\ell$ with $\ell\mid p^2+1$, $\ell\nmid p(p^2-1)$, and $\ord_\ell(p)=4$.
\end{lemma}

\begin{proof}
For $p=2$, take $\ell=5$.
For odd $p$, $p^2+1\equiv 2\pmod 4$ and $p^2+1>2$, so $p^2+1$ has an odd prime divisor $\ell$.
Since $\ell$ is odd and $\gcd(p^2+1,p^2-1)\mid 2$, we get $\ell\nmid p^2-1$, and clearly $\ell\neq p$.
Then $p^4\equiv 1$ and $p^2\not\equiv 1\pmod\ell$, so $\ord_\ell(p)=4$.
\end{proof}

\begin{lemma}[$\ell$-primary automorphisms]\label{lem:coord}
If $u\in\MonSp_n(p)$ is $\ell$-primary, then $u(F)=F$ for some coordinate Lagrangian $F$.
\end{lemma}

\begin{proof}
The permutation $\pi$ induced by $u$ has order dividing that of $u$, which is a power of $\ell$.
For a cycle $(i_1\cdots i_r)$ of $\pi$, including fixed points with $r=1$, let $A_t:E_{i_t}\to E_{i_{t+1}}$ for $t=1,\ldots,r$ be the maps induced by $u$, with indices modulo $r$.
The composite map $A_r\cdots A_1$ is the restriction of $u^r$ to $E_{i_1}$, so it lies in $\Sp(2,p)$ and has $\ell$-power order.
Since $\ell\nmid\lvert\Sp(2,p)\rvert$, it is the identity.
Choose any line $F_{i_1}\subset E_{i_1}$ and set $F_{i_{t+1}}=A_t F_{i_t}$ for $t=1,\ldots,r-1$.
Then $A_rF_{i_r}=F_{i_1}$, so the construction closes consistently, and repeating it on every cycle gives $F=\bigoplus_i F_i$ with $u(F)=F$.
\end{proof}

\begin{lemma}[Descent]\label{lem:descent}
For a coordinate Lagrangian $F$, the space $\Lambda_F(L)=((F\cap L^\perp)+L)/L$ is a Lagrangian of $\Vlog$.
If $u(L)=L$ and $u(F)=F$, then $\rho_L(u)$ preserves $\Lambda_F(L)$.
\end{lemma}

\begin{proof}
The space $\Lambda_F(L)$ is isotropic because $F$ and $L$ are isotropic and $F\cap L^\perp$ pairs trivially with $L$.
Choose complementary lines $F_i^*\subset E_i$ and set $F^*=\bigoplus_iF_i^*$, so that $F$ and $F^*$ pair perfectly.
Let $a=\dim(L\cap F)$.
Projecting $L$ onto $F^*$ along $F$ has kernel $L\cap F$ and an image $P$ of dimension $n-k-a$.
Then $F\cap L^\perp$ equals $\Ann_F(P)=\{f\in F:\langle f,q\rangle=0\text{ for all }q\in P\}$, which has dimension $k+a$.
Hence $\dim\Lambda_F(L)=\dim\bigl((F\cap L^\perp)/(F\cap L)\bigr)=k$, so the isotropic space $\Lambda_F(L)$ is a Lagrangian.
The invariance statement holds because $u$ preserves $L$, $L^\perp$, and $F$.
\end{proof}

\begin{lemma}[No invariant Lagrangian]\label{lem:nolag}
For $k\geq 2$, $\Sp(\Vlog)$ contains an element of order $\ell$ that preserves no Lagrangian.
\end{lemma}

\begin{proof}
Since $\ell\mid p^4-1$, Cauchy's theorem gives an element $g_2\in\Sp(4,p)$ of order $\ell$.
As $\ell\neq p$, $g_2$ is semisimple and has an eigenvalue of 
multiplicative order $\ell$, whose degree over $\Fp$ is $\ord_\ell(p)=4$.
Since $g_2$ is a $4 \times 4$ matrix, this implies that the characteristic 
polynomial of $g_2$ is irreducible (being equal to the minimal polynomial 
of this eigenvalue), and hence $g_2$ has no nonzero proper invariant subspace.
Pick a four-dimensional symplectic subspace $R\subseteq\Vlog$ and set $g=g_2\oplus\Id_{R^\perp}$.
Suppose that $g$ preserves a Lagrangian $\Lambda$.
The subspace $\Lambda\cap R$ is invariant, hence $0$ or $R$, and it is isotropic, so it is $0$.
For $\lambda=r+w\in\Lambda$ with $r\in R$ and $w\in R^\perp$, the vector $g\lambda-\lambda=(g_2-\Id)r$ lies in $\Lambda\cap R=0$.
Since $g_2$ fixes no nonzero vector, $r=0$ and $\Lambda\subseteq R^\perp$.
However, isotropic subspaces of the $(2k-4)$-dimensional space $R^\perp$ have dimension at most $k-2<k=\dim\Lambda$.
\end{proof}

\begin{proof}[Proof of Theorem~\ref{thm:mon}]
Suppose that $\rho_L(\Autmon(L))=\Sp(\Vlog)$.
Take $g$ from Lemma~\ref{lem:nolag} and $u\in\Autmon(L)$ with $\rho_L(u)=g$.
Write $\ord u=\ell^a m$ with $\ell\nmid m$, and set $v=u^m$.
Then $v\in\Autmon(L)$ is $\ell$-primary, and $\rho_L(v)=g^m$ generates the same cyclic group as $g$, so it preserves no Lagrangian.
Lemmas~\ref{lem:coord} and \ref{lem:descent}, however, give a Lagrangian preserved by $\rho_L(v)$, which is a contradiction.
\end{proof}

\section*{Discussion}
Our results show that, in every prime dimension, cheap Clifford gadgets cannot provide full Clifford control over a code that encodes several logical qudits.
Fault-tolerant architectures for high-rate qudit codes therefore need additional ingredients, such as gadgets that use several partitions, code switching~\cite{paetznickUniversalFaultTolerantQuantum2013}, or lattice surgery~\cite{horsmanSurfaceCodeQuantum2012}.
Several questions remain open.
First, for odd $p$, what is the largest subgroup of $\Sp(2k,p)$ realizable by code automorphisms, in analogy with the qubit result of Ref.~\cite{kohAchievingLimitsAutomorphism2026}?
Second, do the theorems extend to Galois qudits of prime-power dimension, where related results are known for phantom codes~\cite{morrisConstraintsPhantomCodes2026} and expected for Clifford circuits on CSS codes~\cite{albertTransversalityStructureClifford2026}?
Finally, can non-Clifford physical gadgets reach logical Clifford gates that Clifford gadgets cannot~\cite{chakrabortyNoGoTheoremFault2026}?

\section*{Acknowledgments}
% Funding taken from projects/replica-advantage (PRXQ round 1, 2026-10); its "startup funding from YMSC" was left out.
We thank A. Chakraborty, H. Yan, and Z. Lu for enlightening discussions. This work is supported by the National Natural Science Foundation of China (Grants No.~T2225008 and No.~T24B2002), the Quantum Science and Technology-National Science and Technology Major Project (Grant No.~2021ZD0302203), the Tsinghua University Dushi Program, and the Shanghai Qi Zhi Institute Innovation Program SQZ202318.
% NOTE(authors): APS requires disclosure of substantive AI use. Lemma 1 (Pauli repair) was also written up with AI assistance.
The proofs of Theorems~\ref{thm:block} and~\ref{thm:mon} were developed by the authors.
AI assistants, including Claude (Anthropic), helped construct Proposition~\ref{prop:tight}, suggested Corollary~\ref{cor:address}, and were used to draft and polish the manuscript.
The authors reviewed and verified all content and take full responsibility for this work.

\bibliography{refs}

\end{document}